# Kinetically-Arrested Phase Separation leads to Tunable Domain Structures in Vapor-Deposited Glasses

A T M Mahbub Alahe[a], Thomas J. Ferron[b], Camille E. Bishop[a,*]

[a]Department of Chemical Engineering and Materials Science, Wayne State University, Detroit, MI 48202, United States

[b]Advanced Light Source, Lawrence Berkeley National Laboratory, Berkeley, CA 94720, United States

[*]Corresponding author, camille.bishop@wayne.edu

## Abstract

The characteristic length scale of phase-separated organic thin film blends is a critical structural parameter governing the performance and functionality of organic electronic devices. The arrested morphologies of vapor-deposited organic thin films result from the interplay between thermodynamic driving forces and kinetic constraints during deposition. Here, we aim to isolate the role of kinetic effects in phase separation by varying the deposition rate at a constant substrate temperature for a co-deposited molecular glass blend of N,N′-bis(3-methylphenyl)-N,N′-diphenylbenzidine (TPD) and Disperse Orange 37 (DO37). The dependence of morphology on deposition rate is quantified using power spectral density (PSD) analysis of atomic force microscopy (AFM) images. Two distinct deposition rate-dependent length scales at the surface reveal how deposition kinetics directly influence domain size and film topography. Complementary Resonant Soft X-ray Scattering (RSoXS) measurements indicate that phase separation extends throughout the film thickness. These observations are consistent with the surface equilibration mechanism previously described for homogenous vapor-deposited films, in which enhanced surface mobility allows molecules in the growing film to partially equilibrate into distinct surface-templated states during deposition. In the current work, this mechanism allows the multi-component blend to phase separate and coarsen into a structure with multiple length scales before kinetically arresting to an extent that depends on the deposition rate. The demonstration of finely tunable domain size with deposition rate provides strategies to design new organic electronic devices with desired morphologies.

## Introduction

Control of domain size during phase separation is important for precise design of organic semiconductor devices. For example, in bulk heterojunction (BHJ) organic photovoltaics, the optimal structure of a donor and acceptor blend is intertwined and bicontinuous, increasing the contact area and shortening the path excitons must travel to the interface.[1–3] In organic field-effect transistors (OFETs), phase separation can enhance electrical performance and long-term stability of the active layer by providing a well-defined dielectric–semiconductor interface in the blend.[4] Device efficiency is strongly dictated by the morphology established during phase separation, particularly the characteristic domain length scale corresponding to the center-to-center distance between domains and the surface topography; therefore, precise control over morphological evolution is essential.[5–11]. This control can be achieved by tuning parameters during film synthesis. Spin coating, die casting, chemical vapor deposition and physical vapor deposition are some of the techniques used to synthesize phase-separated organic semiconducting thin films, but with solution casting in particular, it can be difficult to systematically tune domain structure .[12,13]

The morphology of a phase-separated film is determined by both the thermodynamically-favored structure and the kinetics of the components during separation. A binary mixture can spontaneously phase separate via spinodal decomposition by quenching below the critical point into the unstable region of its phase diagram. The resulting morphology contains interconnected structures of two different phases, with free energy minimization causing it to coarsen during the late stages of separation. On the other hand, if the system is quenched outside the spinodal region but inside the metastable (binodal) region of the phase diagram, phase separation proceeds through nucleation and growth, resulting in a droplet-like structure.[14,15] While composition and temperature determine the thermodynamically allowed phase-separation mechanism[16], the actual pathway in thin films is strongly influenced by kinetic constraints, confinement, and interfacial effects.[17] The interplay of liquid-liquid phase separation (LLPS) and kinetic arrest is a fundamental question with broad applications in materials science, biology, formulations, and the food industry. In protein solutions, the interplay of LLPS with dynamic arrest governs the ultimate phase behavior from what would be predicted by equilibrium theories.[18–22] The interplay between LLPS and slowing kinetics leading to arrest can also be induced by polymerization, such as in the formation of interpenetrating polymer networks.[23,24]

Physical vapor deposition (PVD) can be used to fabricate out-of-equilibrium, yet thermodynamically and kinetically stable organic semiconductor thin films. By changing the substrate temperature and the deposition rate, density, stability, and molecular orientation can be tuned.[25,26] The resulting change in structure can optimize emission efficiency, charge mobility, lifetime, and charge injection in organic electronic devices.[27–29] During PVD, molecules are thermally evaporated in high vacuum, and they deposit on a substrate at controlled temperature and rate. Immediately after molecules deposit on the surface, they experience enhanced mobility. During this brief window, before further deposition buries them and causes an exponential slow-down in dynamics, they can diffuse, rearrange, and even phase separate to a partially (or fully) equilibrated state. This “surface equilibration mechanism” allows molecules to partially equilibrate to a lower energy state with denser and often anisotropic molecular packing.[30] Recent efforts have extended the surface equilibration mechanism framework to structural properties in molecular blends. Co-deposited, evenly mixed blends have higher kinetic stability and density compared to the corresponding liquid-cooled glasses, following similar rules to those established for single-component PVD glasses.[31,32] Additionally, molecular blending can be used to suppress or enhance spontaneous orientation polarization in organic electronic thin films,[33–35] possibly due to nanophase separation.[36] For certain systems, enhanced mobility leads to phase separation and aggregation.[37,38] Recent work has shown that choice of substrate temperature during PVD allow precise control over the size and distribution of phase-separated domains within the film.[39]

In this work, we co-deposited TPD and DO37 thin films via PVD at varying deposition rates and fixed substrate temperature to investigate the kinetics of phase separation in a system with a symmetric (by mass) composition ratio. Co-deposition produces films with well-defined phase separation and a characteristic length scale that can be tuned via changing deposition rate. Higher deposition rates result in smoother films with smaller center-to-center domain spacing, whereas low deposition rates produce larger characteristic length scales and rougher surfaces. The domain evolution of the as-deposited films was studied by further annealing the samples at the deposition temperature, resulting in the evolution of one of the two length scales present in the structure. Surface morphology and length scales were characterized using atomic force microscopy (AFM) and its power spectral density (PSD) to quantitatively determine the characteristic domain spacing. Complementary RSoXS measurements were used to compare bulk phase separation with that observed on the surface, finding a discrepancy in length scales that is undetectable by conventional surface-sensitive characterization methods. We show that phase separation in vapor-deposited blends is tunable using deposition rate only, and that the change in smaller length scales operates independently of large-scale height fluctuations.

## Materials & Methods

Physical Vapor Deposition

N,N'-Bis(3-methylphenyl)-N,N'-diphenylbenzidine (TPD, Tokyo Chemical Industry) and Disperse Orange 37 (DO37, Sigma Aldrich) were co-vapor deposited on both <1 0 0> silicon substrates with native oxide (~2 nm) (University Wafer) and Silicon Nitride Membranes (NX5200C, Norcada) in the same deposition in a vacuum chamber (HEX system, Korvus Technology) with base pressure of $10^{-5}$ mbar. Two crucibles, one with TPD and another with DO37, were heated for evaporation. The rate was measured with a built-in quartz crystal microbalance (QCM). Before actual deposition, the rate of individual molecules was adjusted while the sample shutter was closed. Once both molecules reached the desired rate, the shutter was opened to deposit on the substrate. Deposition was performed on a rotating (20 rpm) substrate at a fixed temperature of 310K at rates from 0.008 nm/s to 0.8 nm/s.

Film thicknesses were measured on the Si substrates only using a spectroscopic ellipsometer (M-2000U, J.A. Woollam) with six incidence angles (45°, 50°, 55°, 60°, 65° and 70°) at ambient conditions and found to be between 72 and 85 nm when fit with a uniaxial anisotropic transparent Cauchy model from (650 – 1000 nm). When not being measured, samples were stored at 248 K on a bed of desiccant.

Atomic Force Microscopy

The height and phase for TPD/DO37 films on Si wafers were measured using a Multimode (Digital Instrument) AFM in tapping mode with Nanoscope V (Bruker) controller and a cantilever (RTESPA300, Bruker) of antimony-doped Si with a spring constant $k$ = 40 N/m and resonant frequency $f_o$ = 300 kHz. All measurements were made in ambient conditions.

The AFM height images were analyzed to quantify surface roughness through two metrics – peak-to-valley height (PV) and root-mean-square roughness. Line-cuts for each deposition rate were used to calculate PV heights as a function of deposition rate. For each sample, $n$=30 evenly spaced 4 µm horizontal linecuts were extracted across the AFM image. The PV height of each linecut $l$ was calculated as

$$PV_l = max\left(h_l(x)\right) - min\left(h_l(x)\right)$$

(Equation 1)

where $h_l(x)$represents the height along linecut $l$ as a function of lateral position $x$. The average PV for each sample was computed as

$$\langle PV \rangle = \frac{1}{n}\sum_{l=1}^{n} PV_l$$

(Equation 2)

and the standard deviation across the linecuts, used as an error estimate, was

$$\sigma_{PV} = \sqrt{\frac{1}{n-1}\sum_{l=1}^{n}\left(PV_l - \langle PV \rangle\right)^2}$$

(Equation 3)

The root-mean-square (RMS) roughness, $R_q$, was calculated over multiple length scales from 50 to 4000 nm by

$$R_q = \sqrt{\frac{1}{N}\sum_{i=1}^{N}(h_i - h_{\text{ave}})^2}$$
(Equation 4)

where $N$ is the total number of pixels in the sampled linecut, $h_i$ is the height at pixel $i$, and $h_{\text{ave}}$is the mean height of the linecut at the specified length scale.

AFM files were analyzed using pySPM.[40] The two-dimensional power spectral density (PSD) was calculated from the AFM height and phase images using a fast Fourier transform.[41,42] The surface height distribution $h(x,y)$ was transformed into reciprocal space, and the PSD was obtained from the squared magnitude of the Fourier coefficients,

$$PSD(q_x, q_y) = \frac{1}{A}\left| \iint h(x,y)e^{-i(q_x x + q_y y)}dx\,dy\right|^2$$

(Equation 5)

where $A$ is the scanned surface area and $q_x$and $q_y$ are the in-plane spatial frequencies. The resulting two-dimensional PSD was azimuthally averaged in reciprocal space to obtain the isotropic PSD as a function of the radial spatial frequency $q = \sqrt{q_x^2 + q_y^2}$. Raw AFM files and code used for processing are included in an online Github repository. Peak fitting was performed by subtracting a linear background and fitting the background-subtracted data to a sum of two Gaussian contributions, with optimization performed with the "curve_fit" function from SciPy's optimize package, as shown in SI Section 1.

Resonant Soft X-ray Scattering

RSoXS was performed at beamline 11.0.1.2 of the Advanced Light Source at Lawrence Berkeley National Laboratory.[43] The sample-to-detector distance was approximately 125 mm and was calibrated using 300 nm polystyrene nanospheres. Two-dimensional scattering was collected on a CCD detector (Teledyne, PI-MTE3) at X-ray photon energies from 250 to 350 eV. All data was reduced and analyzed using PyHyperScattering,[44] with all code and representative scattering data included on Github. Data was dark-subtracted and corrected for exposure time; all films are assumed to be 78 nm thick for normalization. Data was double normalized using the drain current from an in-line gold mesh compared to a reference intensity scan measured on a downstream photodiode (OPTO Diode, SXUV100). Scattering was corrected for photon absorption by dividing by the transmitted flux through the sample measured on the same downstream photodiode. Full data normalization and absorption correction are shown in SI Section 2.

RSoXS peak fitting was done using SciPy.[45] For the high deposition rate data, peaks were fit using the optimize.curve_fit function to a combination of three Gaussians and a linear background; errors from the covariance matrix were minimized and Gaussian fits were visually verified to be physically meaningful, as shown in SI Section 1. Compared to the AFM PSDs, the RSoXS fitting required a third Gaussian to accommodate background scattering at q values less than 0.01 $nm^{-1}$ likely derived from micron-scale surface fluctuations. Due to significantly overlapping peaks and/or extra features from unknown mechanisms, the same fitting procedure did not work for the $R_{dep}$ = 0.04 nm/s data. For these, the peak positions were found using scipy.signal.find_peaks, which identifies local maxima.

The integrated scattering intensity (ISI) was calculated from the reduced one-dimensional RSoXS as

$$ISI = \int_{q_{min}}^{q_{max}} I(q)q^2dq$$

(Equation 6)

In which $q_{min}$ and $q_{max}$ are the lower and upper bounds of the ISI region of interest. Modeling the ISI was performed using a weighted sum of scattering contrasts

$$ISI(E) = B(A|\Delta n_{mat}(E-E_0)|^2 + (1-A)|\Delta n_{vac}(E-E_0)|^2)$$

with three open fit parameters: $B$, a multiplicative scale, $A$, a weighting factor between contrast functions, $\Delta n_x(E)$, the calculated contrast function, and $E_0$, an energy offset to align energy calibrations between reference spectra and current measurements. Fit optimizations were performed with differential_evolution function in the scipy.optimize module. Contrast functions were calculating from near-edge X-ray absorption fine structure (NEXAFS) spectroscopy from prior published work[39] using a Kramers-Kronig transformation (kkcalc[46]) to convert between real and imaginary components of $n(E)$. TPD and DO37 densities were updated according to literature values.[47] In this work, we define the contrast functions with subscripts $vac$, indicating contrast between vacuum and a full-film, volume-average index of refraction of TPD and DO37, and $mat$, representing scattering contrast calculated directly between TPD and DO37.

### Post-deposition Annealing

Samples with deposition rates of 0.008, 0.08, and 0.8 nm/s were annealed at 310 K in vacuum (Linkam, THMSEL350V) for two and half hours. After annealing, samples were cooled back to room temperature at 10 K/min. AFM was performed again to investigate the evolution of surface morphologies and phases. To ensure complete transformation, a second annealing step using the same procedure was performed for selected samples.

## Results

### AFM Surface Morphology and Phase Imaging

TPD, an organic semiconductor molecule, was co-deposited with the DO37 molecule (structures shown Figure 1) at deposition rates of 0.008, 0.04, 0.08, 0.4, and 0.8 nm/s while maintaining a constant substrate temperature of 310 K. All films were prepared with a fixed 50/50 mass ratio of TPD to DO37, equivalent to a 43/57 mole or 54/46 volume ratio. By varying only deposition rate, this results in a systematic investigation of how deposition rate influences film phase separation at identical thermodynamic and composition conditions.

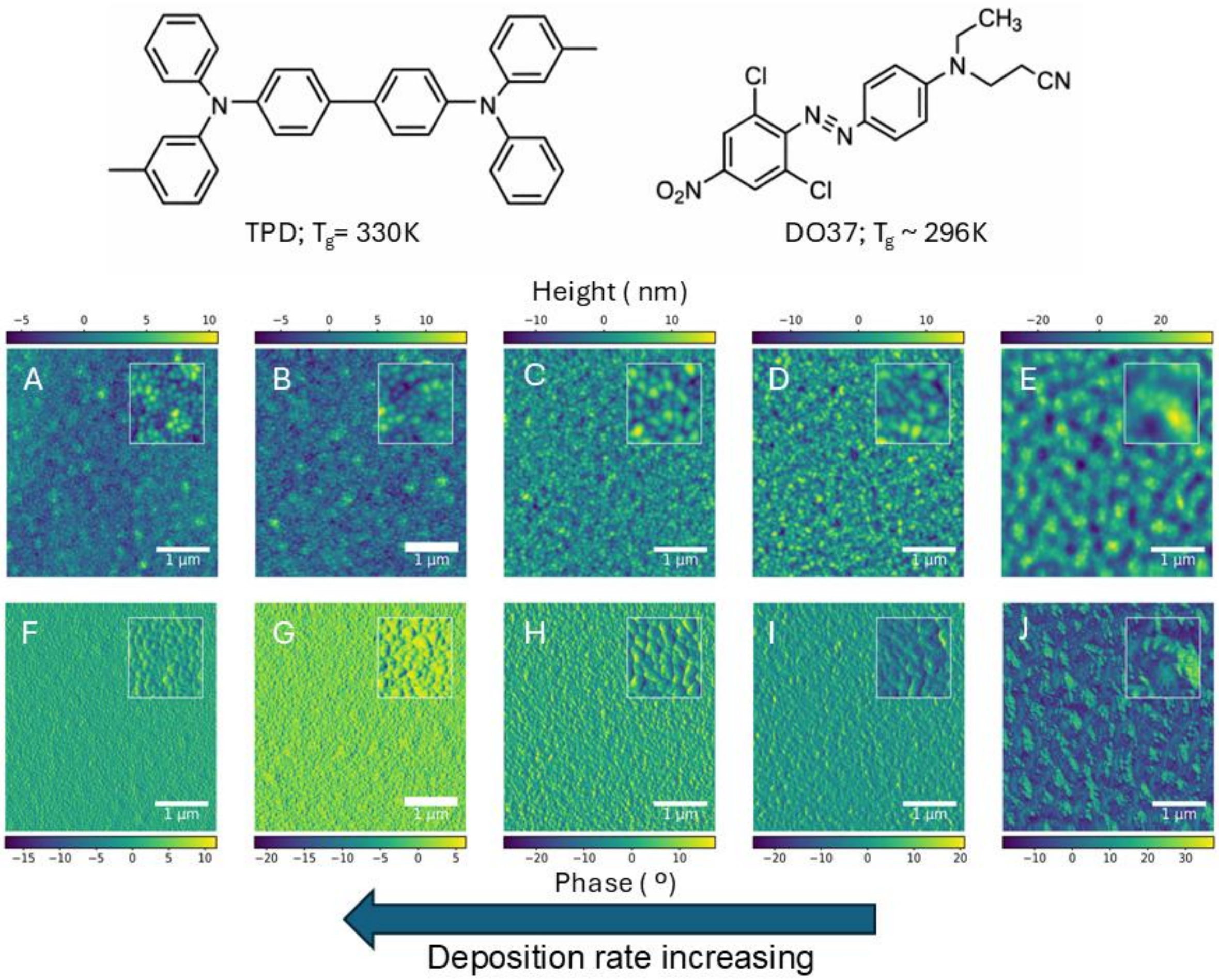

*Figure 1, AFM height (A-E) and phase (F-J) images of samples prepared at deposition rate $R_{dep}$ =: 0.8, 0.4, 0.08, 0.04, 0.008 nm /s from left to right, respectively. All images shown are included at larger resolution in SI Section 3. The maxima and minima of each scale bar vary to maximize contrast in the image. Insets show magnified data from the images they are within.*

AFM height images suggest multi-scale structure formation that changes with deposition rate. At higher deposition rates of 0.8 and 0.4 nm/s (Figures 1A and 1B), the height images show two distinct structures. On a smaller scale, there are nearly circular droplets, while on a larger length scale, clusters of droplets are found with measurable height differences across the surface. As the deposition rate is decreased to 0.08 and 0.04 nm/s (Figures 1C and 1D), the smaller droplets become more irregularly shaped, and the entire film seems to have more uniform variations in height. Finally, when deposited at the slowest rate of 0.008 nm/s (Figure 1E), the smaller features appear less well-defined and a spinodal-like pattern appears in the large-scale height fluctuations. This lowest deposition rate-sample appears qualitatively different from the rest of the higher deposition rates; a similar structure has previously been reported in ultrathin (<50 nm) films of neat TPD on Silicon and on thicker films of TPD on soft substrates.[48,49] In addition to the more extreme height fluctuations, the phase image (Figure 1J) shows notably different behavior from the phase images for the higher deposition rate samples shown in Figures 1F through 1I, which all appear to show a regular, well-defined pattern of droplet-like structures. Due to increased mobility at the surface, lower deposition rates should result in each molecule, on average, spending more time near the free surface with enhanced mobility. This suggests that as the molecules spend an increasing amount of time at the free surface during deposition, they assemble into more cooperatively-formed structures.

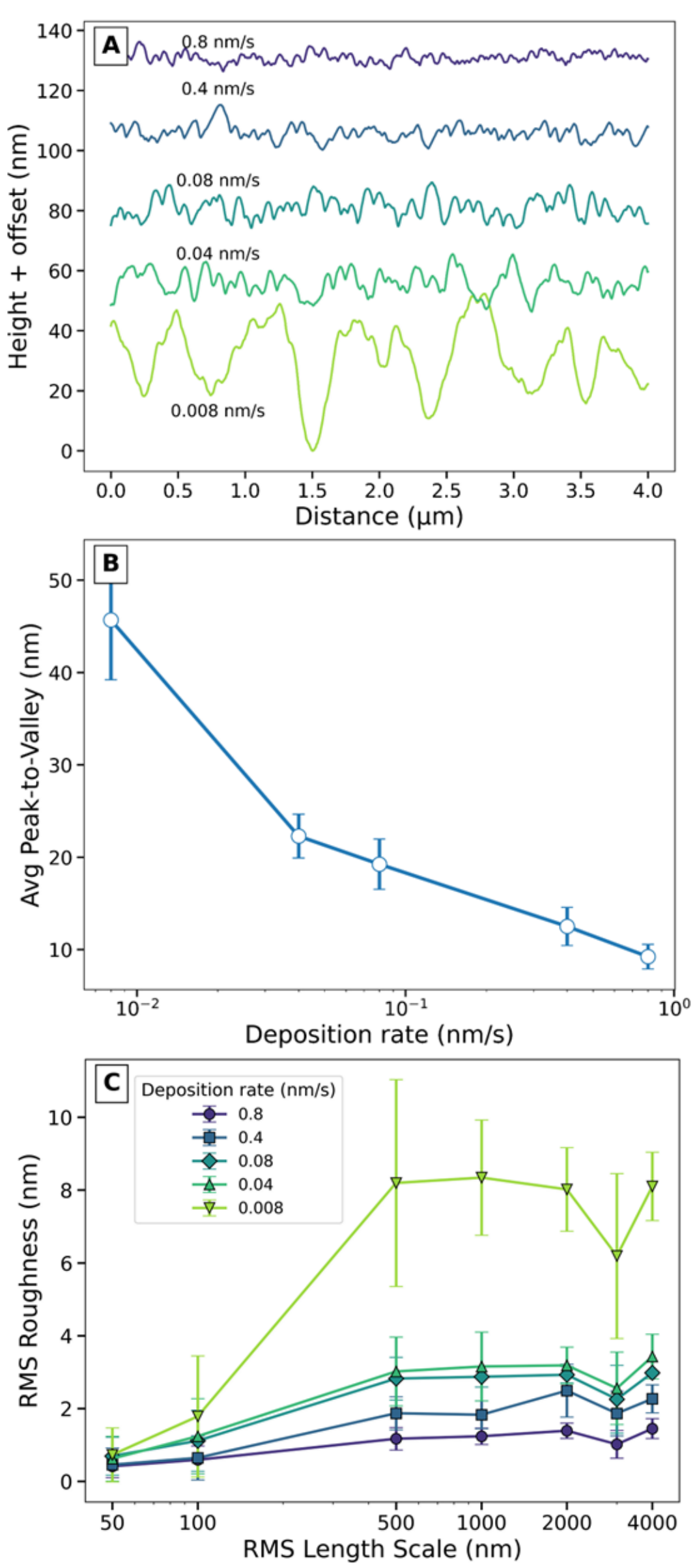

*Figure 2. A) AFM Height linecuts across a single line from each AFM. Profiles are offset by a vertical shift for clarity. B) Average peak-to-valley distance for each sample as a function of deposition rate as calculated by Equation 2, with error bars representing the standard deviation for all linecuts. C) Calculated RMS roughness $R_q$ from equation 4 as a function of linecut length, with the standard deviation from all linecuts represented by error bars.*

When prepared at higher deposition rates, the films are smoother, as seen from the linecuts of the height images shown in Figure 2A. As the deposition rate is lowered, the height fluctuations become dominated by larger length scale features. The quantified peak-to-valley distances (PVs), shown in Figure 2B, obey a smooth logarithmic increase as the deposition rate is lowered, until the rate is lowered to 0.008 nm/s where there is an apparent crossover behavior. The growing roughness at slower deposition rates implies that immediately after deposition, the film is rather smooth, and it roughens as each molecule spends more time at the free surface. Importantly, for the films deposited at the four higher deposition rates, the PV is no more than 30% of the overall ~75 nm film thickness, while the lowest has height fluctuations approximately 60% of the film thickness. Overall, the PV distances

Surface roughness at different deposition rates can be further studied using the root-mean-square (RMS, $R_q$) roughness calculated at different length scales as shown in Figure 2C. This metric quantifies the magnitude of height fluctuations relative to the average surface level at that distance. When $R_q$ is computed over different sampling lengths or window sizes, it enables discrimination between short-scale and long-scale roughness corresponding to local height variations and broader morphological undulations, respectively. By systematically varying the summation length, one can identify the characteristic spatial scales that contribute to the observed surface texture. $R_q$ values are similar for all films when calculated at 50 and 100 nm distances, suggesting that all films are fairly smooth on a short length scale. The $R_q$ calculated over larger length scales (>500 nm) is greater when the film is prepared at the lowest deposition rate. Combined with the PV distances, this suggests that the final self-assembly mechanism for the 0.008 nm/s sample is partially or entirely different from that for the faster deposition rate films.

AFM Power Spectral Density Analysis

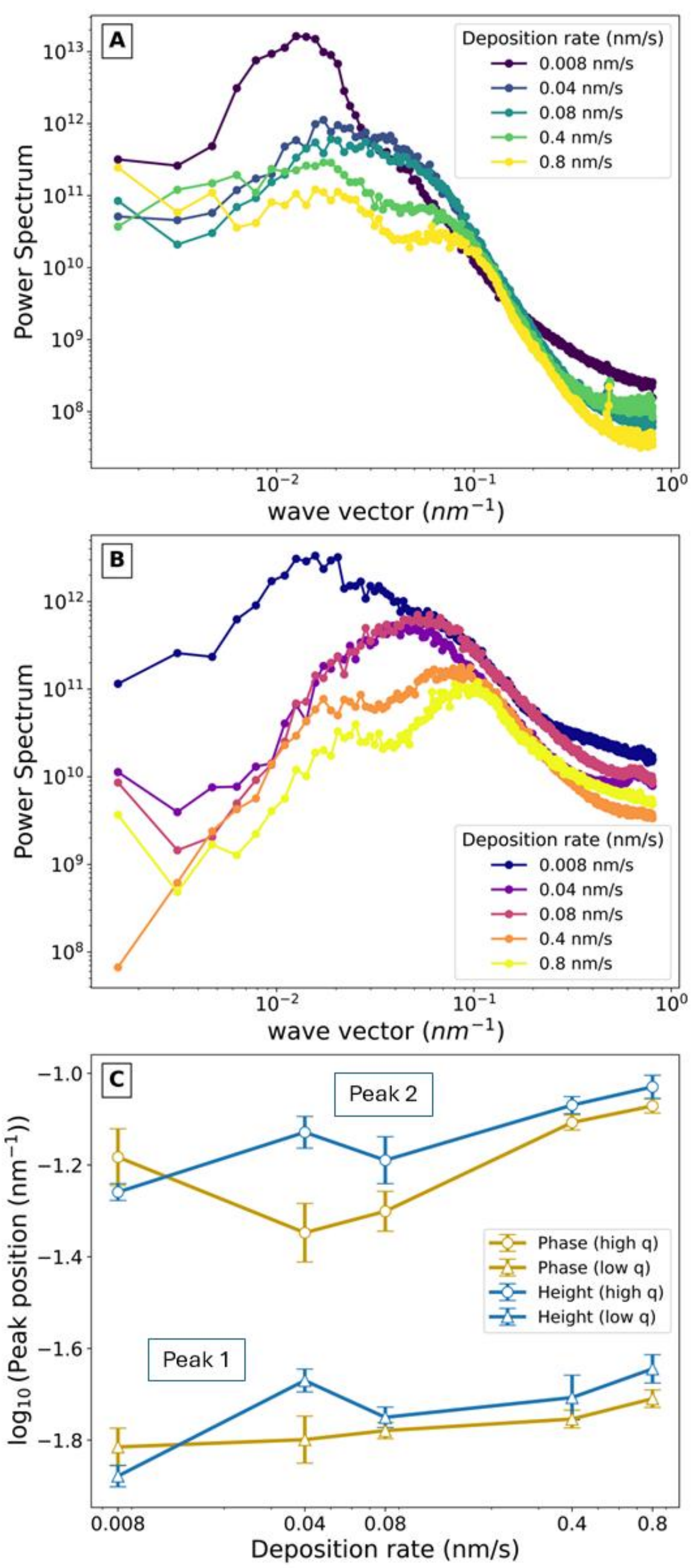

*Figure 3. Fourier-transformed PSDs of AFM A) height and B) phase images for samples at all deposition rates. C) Peak positions in q as a function of deposition rate. "Peak 1" corresponds to the peak at lower q, while "Peak 2" represents the peak at higher q-values.*

Quantitative interpretation of AFM height images shows that length scales can be tuned by changing the deposition rate. In Figure 3, we show the power spectral density (PSD) curves for samples deposited at five rates. Additional PSDs taken at different measurement areas, resolutions, and scan sizes are shown in SI

Section 4. The PSD transforms a two-dimensional real-space image – in this case, corresponding to either height or phase – and plots the intensity vs. the one-dimensional reciprocal space wave vector $q$ in $nm^{-1}$. Peaks in the PSD occur at values of $q$ that correspond to dominant characteristic length scales $d$ that appear in the image, where $d = 2\pi/q$. These peaks represent the structure factor of height and possibly composition fluctuations in the material, and are frequently employed to measure phase separation from real-space imaging.[50]

For the height and phase PSDs shown in Figures 3A and 3B, two distinct length scales are present at the four highest deposition rates, with a clear shift in behavior for the lowest $R_{dep}$ sample (0.008 nm/s). As opposed to the two qualitatively similar peaks present in each higher $R_{dep}$ sample, the 0.008 nm/s film has a clear peak at q ~ 0.013 $nm^{-1}$ and only a weak shoulder near q ~ 0.055 $nm^{-1}$. This weak shoulder marks a distinct break in the trend observed in the higher $R_{dep}$ samples and indicates that structural correlations are not as strong at a smaller length scale. For both the height and phase PSDs, only the two highest $R_{dep}$ samples - 0.4 and 0.8 nm/s - show two clear peaks. Samples deposited at 0.04 and 0.08 nm/s appear to have a single broad feature that, for now, we will claim is the convolution of two length scales with more evidence provided later. Therefore, the uncertainty in peak position is greater for these two lower deposition rates, leading to the larger error bars and possibly the apparent increase in q-position for the high-q peak in the $R_{dep}$ = 0.04 nm/s sample. Peak fitting procedures are shown in SI Section 1.

The two peaks in the PSDs have different dependences on deposition rate, as shown in Figure 3C. For the peak at q-values near q ~ 0.015 $nm^{-1}$, $R_{dep}$ has little to no effect on the peak position. There is a slight decrease in peak position for this peak 1 as the deposition is slowed, which we attribute to time spent on the substrate during deposition; this process will be discussed later in the context of an annealing experiment shown in Figure 6. In contrast, peaks at higher q-values (peak 2) show a strong dependence on $R_{dep}$, shifting to lower values of q (larger real-space length scales) at lower deposition rates. While the peaks in the height and phase PSD depend similarly on deposition rate, the two are offset from one another by a small constant offset, as shown in Figure 3C. The PSDs were calculated for the height and phase images from a single scan. It is not clear whether the difference in peak position is due to a clear physical difference between the height and the phase image, or simply an artefact of one of the two AFM channels, so we do not attempt to assign a physical origin to this disagreement.

Bulk structure analysis with Resonant Soft X-ray Scattering

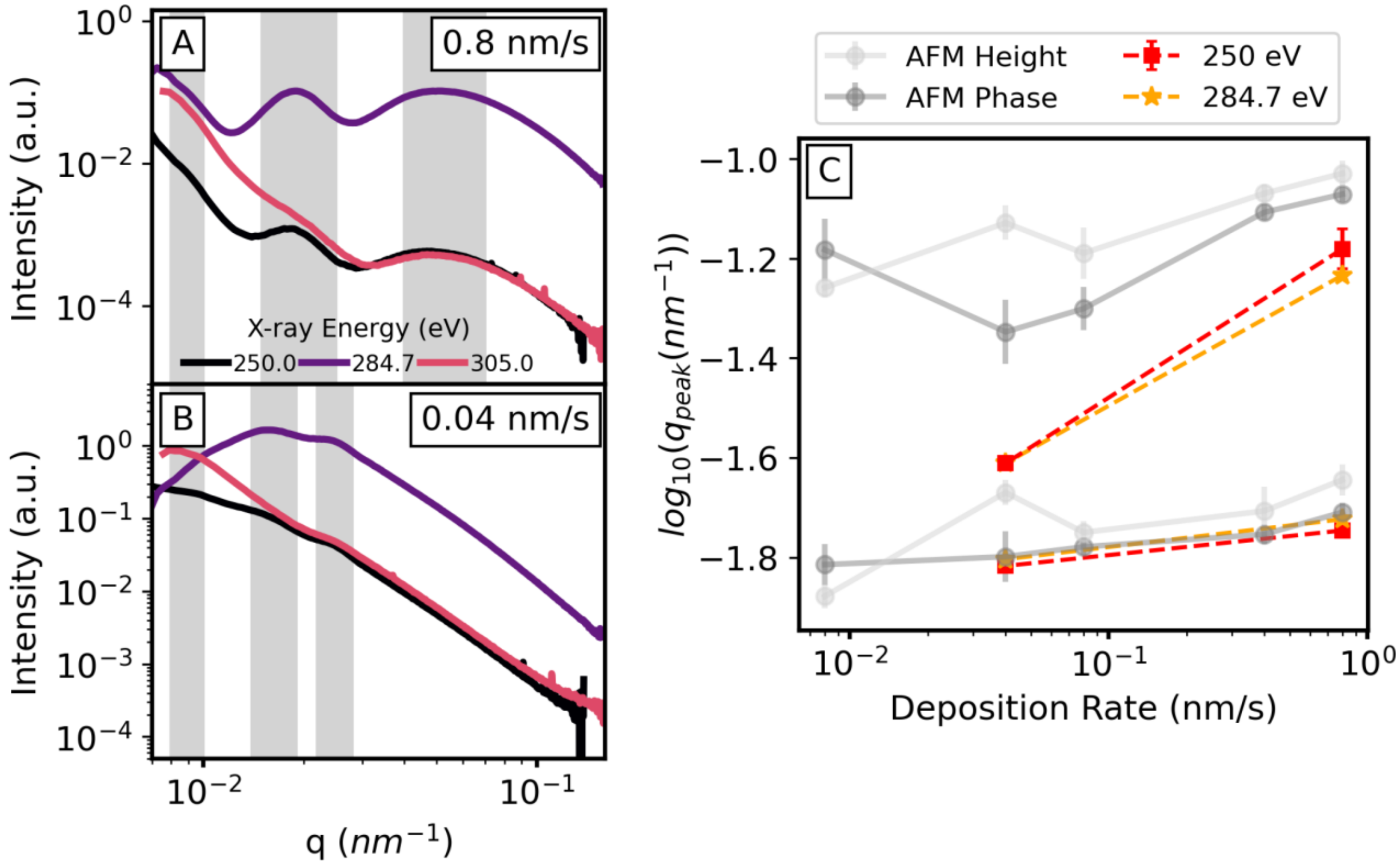

*Figure 4. Isotropic resonant soft X-ray scattering (RSoXS) intensity vs. scattering vector q of blends prepared at a deposition rate of A) 0.8 and B) 0.04 nm/s. Shaded gray regions correspond to regions of integration in later analysis. C) q-positions of the RSoXS peaks (colored symbols) at two photon energies compared to the peak positions from AFM PSDs (grey, also shown in Figure 3).*

Since the AFM measures only the top few nanometers of the film, we use transmission RSoXS to measure the bulk morphology due to its enhanced contrast between primarily carbon-containing domains. In Figures 4A and 4B, we show the azimuthally-averaged RSoXS scattering intensity as a function of the scattering vector $q$ for samples deposited at 0.8 and 0.04 nm/s, respectively. We have verified that there is no polarization-induced anisotropic scattering as data measured from orthogonal polarizations are identical, as shown in SI Section 5.[51] All analysis will only be conducted using horizontally polarized X-rays.

Samples deposited at both rates produce two Bragg-like peaks consistent with the corresponding PSDs. These peaks are not at spacings that would correspond to harmonic reflections suggesting that there is significant phase-separation across two well-defined length scales that propagate throughout the film. Furthermore, grazing-incidence wide-angle X-ray scattering (SI section 6) confirms that these films are amorphous and thus this scattering does not originate from buried crystalline phases. This result is similar to the scattering seen in a prior publication on TPD/DO37[39], but distinct from that seen in a co-deposited system of TPD/TCTA, which shows only form factor scattering.[52] Similarly to the PSDs shown in Figure 3, when prepared at a lower deposition rate, the two peaks in the resulting scattering pattern merge together. Interestingly, peak 2 (higher q) measured by RSoXS is at a much lower q-vector than found by AFM, while peak 1 is at a nearly identical position compared between the two measurements. As shown in Figure 4C, for the lower deposition rate sample, peak 2 occurs at 0.025 nm$^{-1}$ when measured by RSoXS, but it appears at 0.045 nm$^{-1}$ when measured by AFM. This corresponds to a discrepancy in measured real-space distances of $d_{RSoXS} = 250\,nm$ to $d_{AFM} = 140\,nm$ by RSoXS and AFM, respectively. Similarly, for the higher deposition rate sample, peak 2 is at 0.045 $nm^{-1}$ by RSoXS and 0.089 $nm^{-1}$ by AFM, corresponding to a disagreement in length scales of 140 nm to 70 nm. Discounting any miscalibration in the scattering geometry, which would impact both peak positions equally, this suggests that the buried structure is not the same as that observed by AFM; we comment further in the Discussion section.

The intensities of the peaks vary strongly with energy, with I vs. q patterns taken at 250.0 eV and 305.0 eV displaying only subtle peaks compared to 284.7 eV which is near resonance. Data for all 114 photon energies measured is included in SI Section 7. Generally, at X-ray energies well below the carbon K-edge (284 eV), scattering signal from voids and/or surface roughness is enhanced relative to material domains.[53] Near the absorption edge, where compositional contrast between the two organic components is maximized, the peaks become the most pronounced if they originate from composition separation. Measurements made at the non-resonant, pre-edge 250.0 eV (red squares) yield, within error, nearly identical peak positions to those made at the resonant, on-edge 285.2 eV (orange stars). The consistency in peak position with energy suggests that the phase separation and surface topography in this sample are highly correlated; we further explore the contrast mechanisms by quantifying the scattering intensity as a function of measurement energy.

The variation of scattering intensity with respect to photon energy can be used to infer the physical origin of scattering features. This behavior is analogous to X-ray spectroscopy where the integrated scattering intensity (ISI) over a fixed q-range is proportional to a sum of binary contrast functions, $C(E) = |\Delta n(E)|^2$, weighted by the relative volume of scattering phases. Contrast is calculated using the difference in complex index of refraction, $n(E)$ where $E$ is the photon energy, of components such that $n(E) = 1 - \delta(E) + i\beta(E)$ where $\delta(E)$ is related to phase scattering and $\beta(E)$ is proportional to photon absorption. Complete calculation of binary contrast functions is shown in SI Section 8.

The solid black line in Figure 5A shows the calculated contrast due entirely to compositional differences between TPD and DO37. Binary contrast between two resonant materials typically shows a positive pre-edge slope before reaching a maximum and subsequent fine structure, which TPD and DO37 follow up to the first maximum around 284.9 eV. In contrast, a resonant and non-resonant contrast function (e.g., material vs. vacuum) decreases toward a minimum before a sharp increase at the contrast maximum. The lightest grey line, representing contrast between vacuum ($n(E) = 1.0$) and a homogeneous blend of TPD and DO37, shows a minimum near 283 eV before exhibiting a similar behavior above the absorption edge. Resonant fine-structure at and above the Carbon K-edge is highly sensitive to the molecular conformation and orientation of the molecules under investigation. For molecules such as DO37, which can adopt both *trans* and *cis* configurations depending on PVD conditions,[54] it is difficult to identify a quantitative reference index of refraction. The reference spectrum used in this study likely contains a population of both configurations which may differ from the blended films in the current study. However, the pre-edge behavior is universal as it originates from a near contrast-match between the indices of the material and vacuum, and therefore can reliably distinguish whether scattering predominantly comes from material domains or vacuum correlations.

The ISI allows us to isolate the scattering intensity at each length scale to determine the degree to which different features are caused by surface roughness, compositional separation, or both. We assume that these films do not contain internal voids and therefore attribute all "vacuum" scattering to surface roughness. We proceed with this analysis investigating three features of interest: the Bragg-like peaks that are similarly found in the AFM data, referred to as peak 1 and peak 2, and the broad shoulder at low scattering vector, referred to as Low q. The ISI for each feature of interest is shown in Figure 5 as colored symbols. Integration limits for each region are shown in Table 1 and represented by the grey shaded boxes in Figures 4A and 4B. Fits between 270 and 284 eV are shown in greyscale in Figures 5B and 5C and are used to estimate the percentage of scattering due to compositional correlations. Fitting is performed with differential evolution optimization as described in Materials & Methods. The full range of ISIs from 250 to 350 eV is shown in SI Section 9.

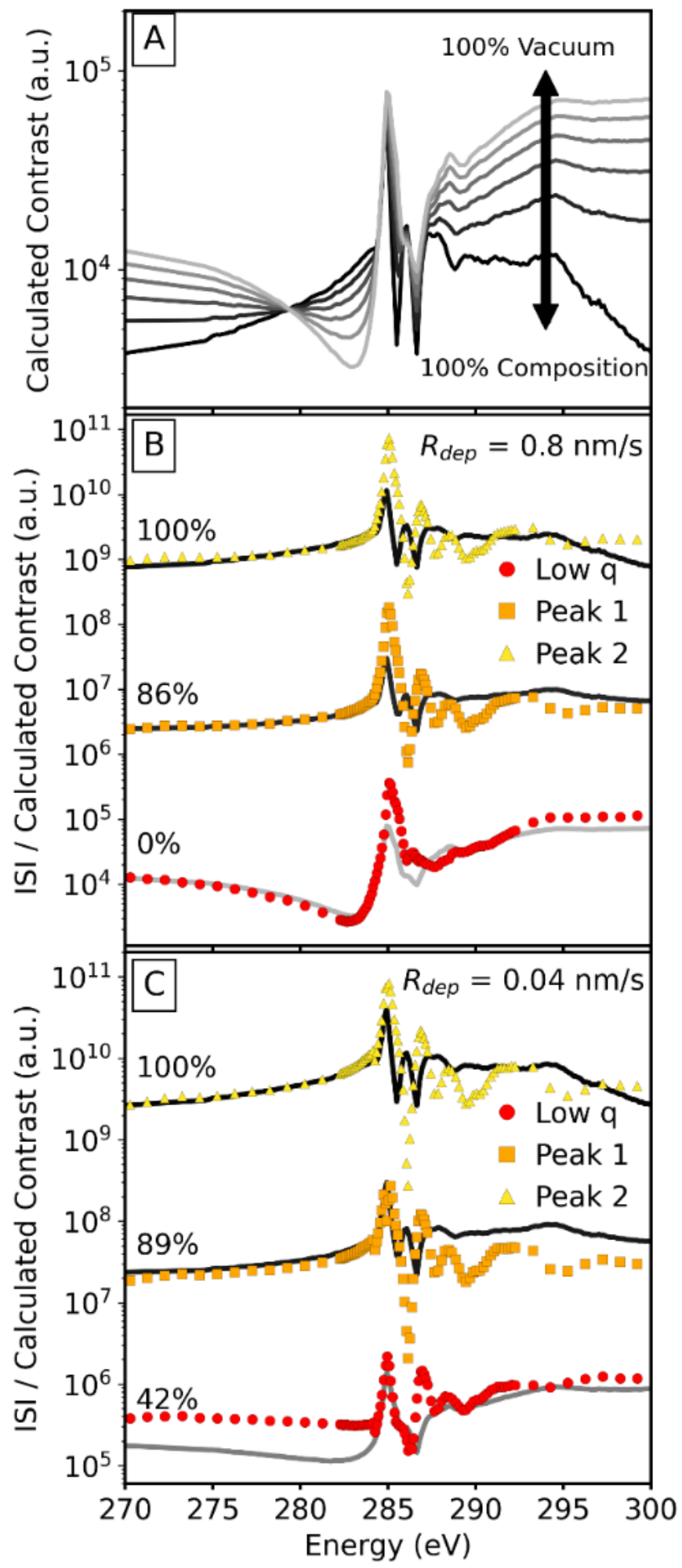

*Figure 5. Comparison of experimental integrated scattering intensity to calculated scattering contrasts for material and vacuum. A) Calculated scattering contrasts as a function of energy for different relative contributions of material-vacuum ("vacuum") and material-material ("composition") scattering mechanisms. B) and C) Experimental ISI (colored symbols) for the two deposition rates overlaid on best fits of calculated scattering contrast (greyscale). All data is in arbitrary units and vertically shifted for clarity. Percentage contribution of composition scattering (fit parameter, A, described in Materials & Methods) is given above each ISI.*

The calculated ISI indicates that the Bragg-like peaks in the RSoXS data have a different origin than the low-q shoulder. Our modeling suggests that long length scales ca. 1 μm are primarily driven by height fluctuations, as the ISI fit for the low-q data is best described by $|\Delta n_{vac}|^2$ ($A < 0.5$). Qualitatively, this can be seen by the minima around 283 eV for $R_{dep} = 0.8$ nm/s and the flat pre-edge for $R_{dep} = 0.04$ nm/s. The difference between the two behaviors highlights the overlap with all scattering features in $R_{dep} = 0.04$ nm/s where, while the integration limits do not overlap, the shift in peak position makes their physical overlap greater. These partial contributions suggest that at lower deposition rates, material begins to phase separate at much larger length scales commensurate with height fluctuations.

Models for peak 1 and peak 2 indicate that compositional phase separation is the dominant contribution to the scattering. As shown in Figures 5B and 5C, the energy dependence of $ISI_{Peak\,1}$ for both $R_{dep}$ corresponds to roughly 85-90% composition contrast, and that for $ISI_{Peak\,2}$ corresponds to 100% composition contrast. Qualitatively, this is visible in the suppression of the pre-edge slope for $ISI_{Peak\,1}$ in both samples in Figures 5. This subtle difference indicates that peak 1, the larger length scale, has a greater contribution from the surface topography while peak 2 is more concentrated in the bulk of the sample, which suggests a reason for the disagreement shown in Figure 4C. The topography on the surface has had additional time to "age" compared to the film bulk causing a subtle difference in the characteristic length-scale of the phase-separation.

| $R_{dep}$ (nm/s) | Label | $q_{min}$ ($nm^{-1}$) | $q_{max}$ ($nm^{-1}$) |
|---|---|---|---|
| 0.8 | Low q | 0.08 | 0.10 |
| 0.8 | Peak 1 | 0.15 | 0.25 |
| 0.8 | Peak 2 | 0.40 | 0.70 |
| 0.04 | Low q | 0.08 | 0.10 |
| 0.04 | Peak 1 | 0.14 | 0.19 |
| 0.04 | Peak 2 | 0.22 | 0.28 |

*Table 1. Integration limits for ISIs shown in Figures 5A and 5B.*

Morphology evolution during annealing

To determine whether the films are rearranging in bulk during deposition, the as-deposited films were annealed in vacuum for two hours at the deposition temperature, 310 K, and subsequently measured by AFM to quantify length scale and roughness changes as shown in Figure 6. The two length scales in the as-deposited films evolve independently as they are held longer at the deposition temperature. The small features in the film deposited at the fastest $R_{dep}$ of 0.8 nm/s shown in Figure 6A (the same film as shown in Figure 1E) qualitatively appear to "smooth" out when annealed (6B), resulting in a slightly decreased intensity of peak 2 in the PSD shown in 6C. Despite the decreased intensity, the peak position remains at the same location in q, showing that the short real-space length scale stays constant. The lower q peak 1 in the PSD, however, moves to lower q (larger real-space length scale) until it reaches a limiting value that is similar to that found for the as-deposited 0.008 nm/s film. Nearly identical behavior occurs in the medium $R_{dep}$ = 0.08 nm/s film shown in Figures 6D – 6F, with the previously merged peaks separating as the low q peak moves to even lower q with annealing. The shift of peak 1 with annealing may explain the slight decrease in q-position with slower deposition rate shown in Figure 3C. While the position of peak 1 appears to be roughly constant with deposition rate, slower deposition rate samples stay at the deposition temperature for longer; presumably, this would give the sample time to gradually shift that length scale to the lower q. For the lowest 0.008 nm/s deposition rate film, Figures 6G – 6H show that the two dominant length scales do not change with further annealing. The change in peak position due to annealing is shown in Figure 6J. The difference in behavior for peak 1 and peak 2 with annealing in the two higher deposition rate samples suggest that the physical processes that result in those length scales occur independently of one another. A further two hours of annealing did not show any more evolution in PSDs (SI Section 10).

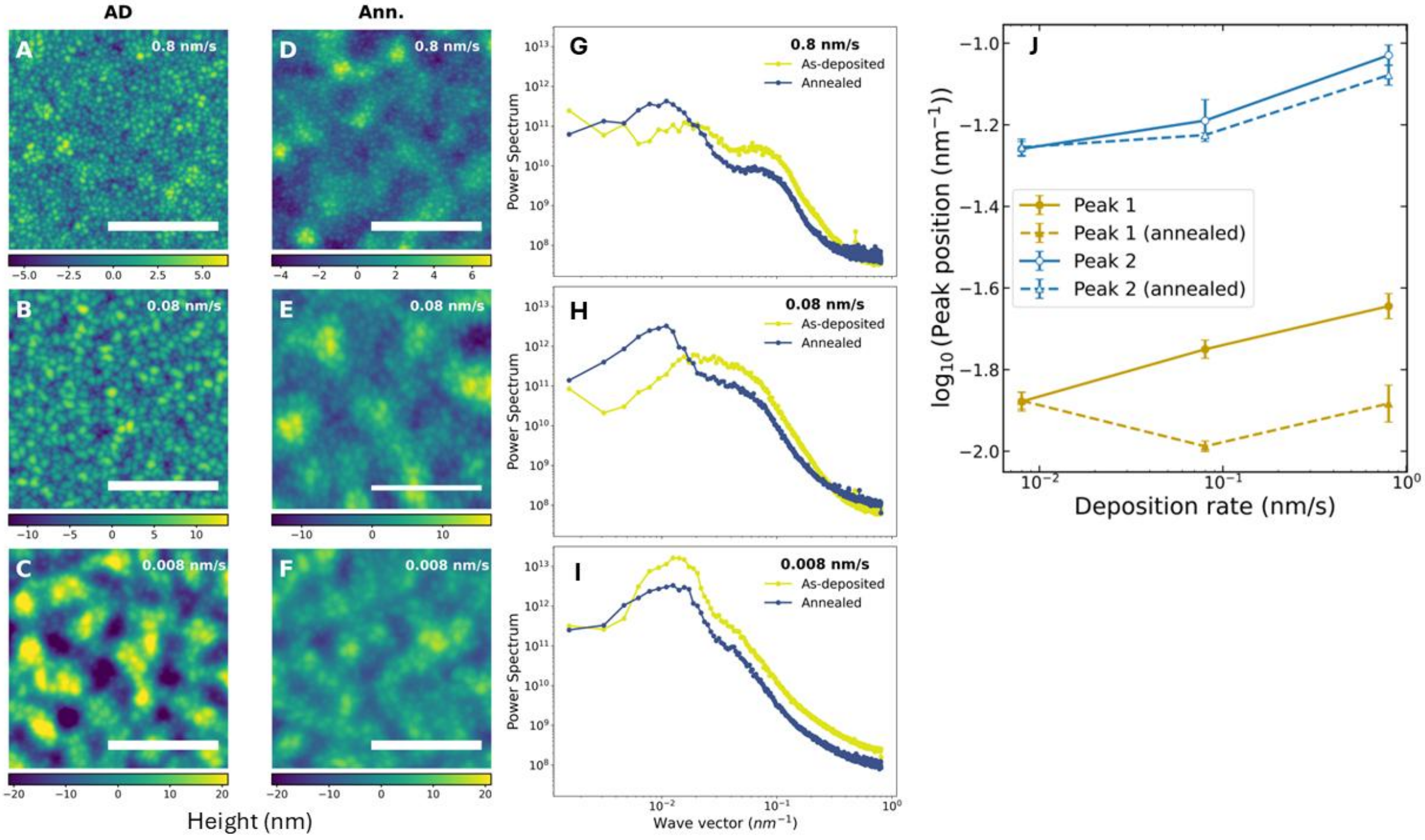

*Figure 6. Evolution of AFM profiles with annealing in vacuum at 310 K (equal to the substrate temperature during deposition). AFM height images of as-deposited samples with $R_{dep}$ equal to A) 0.008, B) 0.08, and C) 0.8 nm/s. Corresponding AFM height images of samples after annealing for 2 hours are shown for the three deposition rates next to their corresponding as-deposited image in D), E), and F). G) – I) PSDs from the height images before and after annealing the samples deposited at the three deposition rates. J) Summary of fit peak positions in q before and after annealing for the three samples. AFM scale bars are 1 µm; as-deposited films are scans from the same samples as shown in Figure 1, but in different areas and at a smaller scan size.*

Conversely, the 0.008 nm/s film appears to flatten with annealing, while the roughness of the two higher deposition rate films stays almost constant. Figure 7A shows horizontal linecuts of the films before and after annealing. For all three samples, local height fluctuations smooth out once the film is annealed. However, on a longer scale, the behavior of the 0.008 nm/s sample is distinct, as it smooths on a larger length scale as well. This trend is reflected in the peak-to-valley distance (PV), shown in Figure 7B. Within error, the PV of the 0.8 and 0.08 nm/s sample do not change with annealing, while that of the slow deposition rate film reduces by nearly 20 nm – 27% of the thickness of a 75-nm film. This is similarly reflected in the RMS data shown in Figure 7C, in which the RMS roughness is computed over different length scales. Upon annealing, the 0.008 nm/s film becomes smoother at all length scales and ends with values similar to that of the film deposited at 0.08 nm/s. Interestingly, the 0.8 nm/s and 0.08 nm/s films do not start with the same values of RMS roughness, and they remain distinct after annealing without changing appreciably. As with the PSDs, the RMSs and PVs did not change with two subsequent hours of annealing (SI Section 10), indicating that they have completely evolved in the above analysis.

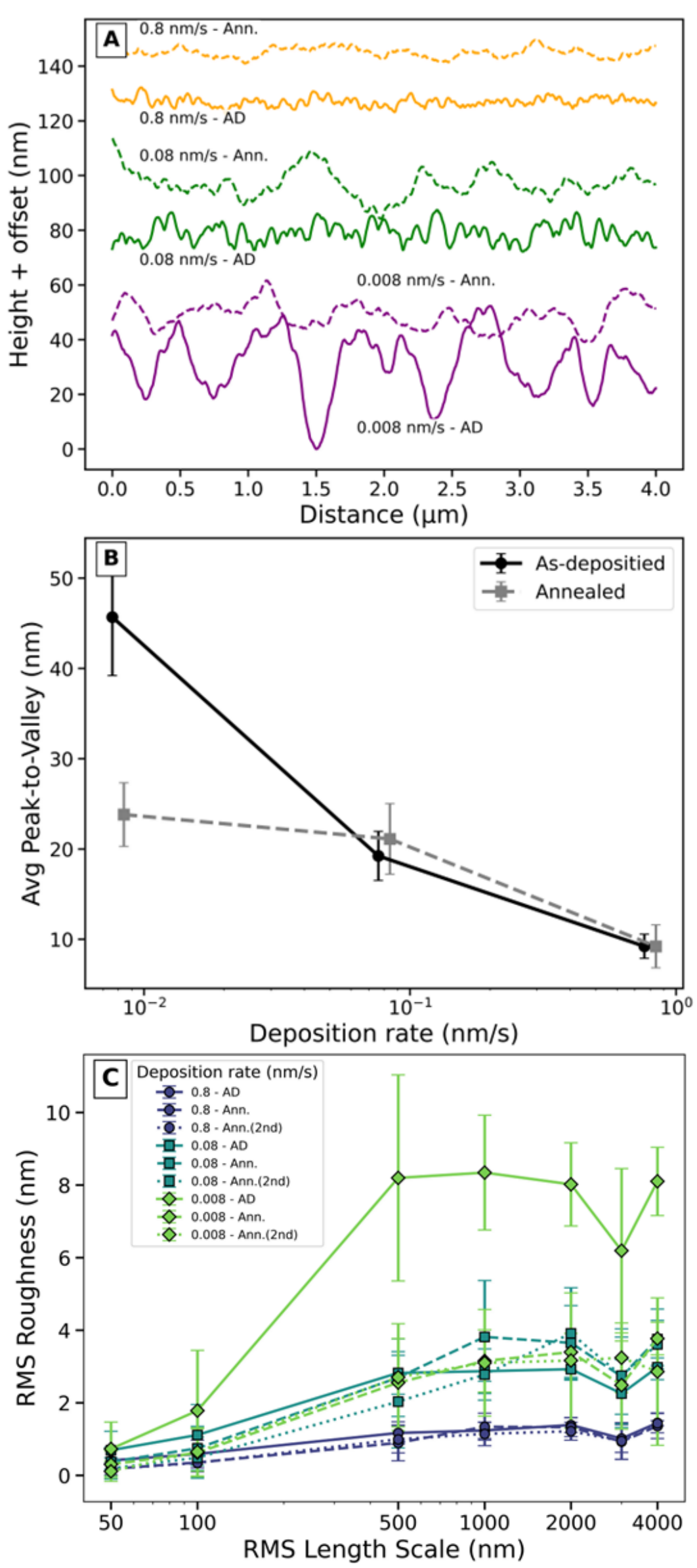

*Figure 7. Quantifying the magnitude of height fluctuations after annealing at 310 K. A) Representative linecuts for the three deposition rates before (solid lines) and after (dashed lines) annealing. Curves are arbitrarily vertically shifted for clarity. B) Calculated peak-to-valley (PV) distances as a function of deposition rate before and after annealing. C) RMS over different length scales for the three deposition rate samples before and after annealing.*

## Discussion

The two distinct length scales observed in both AFM and RSoXS indicate the coexistence of multiple morphological features within the thin film which can be explained by extending the surface equilibration mechanism for vapor-deposited glasses to phase separating systems. Vapor-deposited thin films have enhanced surface mobility which allows surface molecules to partially equilibrate before becoming part of the bulk.[30] The phase separation and coarsening of domains in the present work can be understood whereby the enhanced mobility allows the molecules to partially achieve an underlying equilibrium phase-separated structure at the free surface. In the case of an immiscible system, this mobility allows molecules to phase separate well below the bulk glass transition temperature. Since the surface of a PVD-grown film remains highly mobile during deposition, films grown at different rates undergo phase separation and coarsening to different extents before the structure becomes kinetically arrested. Samples prepared at lower deposition rates have features with larger length scales due to the extended amount of time molecules spend at the surface with enhanced mobility, allowing domains to grow. As a result, the final as-deposited films have distinct morphologies and characteristic length scales that can be predictably tuned by changing the deposition rate, even at a fixed substrate temperature.

The phase separation in this work appears to proceed by a two-step mechanism that results in two distinct kinetically-trapped length scales. Peak 2, at higher q, is attributed to cluster or droplet-like morphologies, reflecting short-range structural correlations supported by PSDs of filtered and binarized images, shown in SI Section 11. In contrast, peak 1 corresponds to features associated with a bicontinuous-like, spinodal structure, indicative of long-range compositional and/or height fluctuations. Such morphological coexistence was first reported by Takeno and Hashimoto in the context of a percolation-to-cluster transition (PCT).[55] In their study of annealed phase-separating polymer systems, the PCT was described as arising from a secondary kinetic crossover during phase separation, whereby an initially bicontinuous, percolated network evolves into discrete cluster or droplet domains. The present observations are consistent with this framework and the surface equilibration mechanism, suggesting that the system follows a similar kinetic pathway boosted by enhanced mobility and kinetic arrest in which percolated and clustered morphologies transiently coexist within the accessible experimental time window.

Both the AFM images and the RSoXS performed in this work are different from previously-published results on TPD/DO37;[39] the comparison between the RSoXS $I\ vs.q$ is plotted in SI Section 12. Qualitatively, the AFM in the prior work clearly shows two distinct phases, whereas in this work, the domains have less well-defined “matrix” and “hole” morphology. The RSoXS in the current work has a high-q peak at a value lower than would be expected from the prior data. While we cannot be entirely sure of the reasons for these discrepancies, we note that films from both works are amorphous and have length scales that can be tuned by either substrate temperature or deposition rate. Phase separation and morphology in vapor-deposited films can be highly sensitive to thickness differences of even 5 nm.[48,56] While our film thicknesses have a range around $79 \pm 6$ nm, the consistent trend in length scales between samples with varied deposition rates suggests that in the current work, we have sufficiently similar film thickness to fairly compare structure between all films.

One surprising result of the RSoXS vs. AFM comparison is that while the large length scales of peak 1 agree between the AFM and RSoXS, the high q length scales differ significantly. In prior work, the AFM- and RSoXS-determined length scales agree;[39] however, for both samples in the present work, the real-space length scale measured by AFM is approximately half as large that measured by RSoXS. We propose the following explanation as one possibility. During deposition, the initial stage of domain formation is through island formation and coarsening. Upon further deposition, material on top of these domains will “fuse” them together into larger domains. As this is the structure present through the bulk of the film, which comprises at least 90% of the sample, it is the dominant length scale measured by RSoXS. Because there is eventually uncapped surface, one “layer” of domains is not as thick and does not fully coarsen, and this is the structure that is measured by AFM. Recent studies on vapor-deposited ultrastable metallic glasses find that alloys grow in a periodic island-layer-island growth, suggesting that there could be two separate stages of growth in our current system.[57]

The domains in the samples prepared in this work are larger laterally than the film thickness, suggesting domains that are wider than they are tall. The annealing experiment shown in Figure 6, in which the higher-q peak 2 position does not evolve with further annealing at the deposition temperature of 310 K, suggests that the contrast between enhanced free surface mobility and hindered bulk mobility are essential for the phase separation behavior displayed. There are two distinct hypotheses – that the separated phases are essentially columns that propagate through the depth of the sample, or that phase separation occurs in distinct layers. The well-defined structure factor apparent in the reciprocal-space data suggests a collective, non-random process. A possibility is that as the film grows, the freshly deposited molecules collectively assemble with the molecules directly beneath them; in this way, the entire film "collectively" phase separates and assembles throughout the deposition process, but does so only a few nm at a time. Prior work suggests that the structure of a vapor-deposited glass depends on the depth of equilibration during deposition and that a lower effective deposition rate results in a film equilibrated to a greater depth – as deep as 9 nm for an effective rate of 0.04 nm/s.[58,59] As the effective deposition rate depends on both true deposition rate and substrate temperature, variable temperature deposition could be used to test this theory.

It is notable that the shorter characteristic length scale (peak 2) remains unchanged with annealing for all three samples, while the large length scale (peak 1) evolves to be the same for all three films. The RMS roughness analysis of the annealed samples revealed that, for the lowest deposition rate of 0.008 nm/s, the global surface roughness decreases upon annealing; similar behavior is observed in previous reports of neat TPD deposited on soft substrates.[49] This suggests that, although lateral domain coarsening at the free surface becomes arrested when the system reaches equilibrium, capillary leveling continues to smooth the surface height fluctuations. This occurs because reduction of surface curvature is a surface tension-driven relaxation process[60] that operates independently of the underlying phase-separated morphology. These observations indicate that phase separation equilibrium is not equivalent to surface mechanical equilibrium, and that capillary leveling selectively smooths height without affecting composition. A coexisting or alternate theory is that the intermolecular interactions within the structures that lead to the high q peak are similar to those that are observed in ultrastable vapor-deposited glasses. Due to the surface equilibration mechanism, vapor-deposited glasses often have enhanced stability and denser packing.[25] This raises the onset temperature of transformation ($T_{onset}$) well above that of the ordinary glass transition temperature $T_g$; therefore, one possibility is that the annealing temperature of 310 K is below $T_{onset}$ for this structural feature of the as-deposited glasses. It is remarkable that this applies to only one length scale of ordering. Additionally, this result suggests that the tunability of the shorter length scale is possible only by physical vapor deposition.

## Conclusions

In summary, we find that the phase separated structure of a vapor-deposited organic glass blend can be tuned by varying only the deposition rate. As molecules spend more time at the free surface at lower deposition rates, they are allowed more time to phase separate before they are kinetically arrested in a non-equilibrium state. Combined surface-sensitive AFM and bulk RSoXS measurements determine that the length scales differ in the bulk and at the surface. An RSoXS scattering intensity analysis shows that smaller length scales in the film are due almost entirely to composition separation, while larger length scales are due either entirely to height fluctuations or to a combination of height and composition fluctuations. Further annealing experiments found that the short length scales remain while the larger length scales continue to shift. The results of this work can be explained by the existing surface equilibration mechanism for vapor-deposited glasses and expand its usefulness to phase separating blends that could be useful for organic electronic applications.

## Supplementary Material

Description: AFM PSD and RSoXS peak fitting; RSoXS normalization and absorbance correction; enlarged AFM images; PSDs of all AFMs; verification of scattering isotropy; grazing-incidence wide-angle X-ray scattering (GIWAXS); additional RSoXS data and example 2-dimensional images; X-ray absorption spectroscopy (NEXAFS) and binary scattering contrast calculations; full measured energy range of ISI in Figure 5; AFM height and phase images and PSDs from annealing experiment; image processing for physical origin of structural correlations; comparison to published work.

## Acknowledgements

A.T.M.M.A. and C.E.B. acknowledge Wayne State University faculty start-up funds as the primary source of funding for this work. This research used Beamline 11.0.1.2 of the Advanced Light Source, a U.S. DOE Office of Science User Facility under contract no. DE-AC02-05CH11231. The authors gratefully acknowledge Nurjahan Khatun and Chinecherem Nkele of Kushal Bagchi's research group for GIWAXS measurements at the Stanford Synchrotron Radiation Lightsource Beamline 11-3. The authors also acknowledge Ali Hatami of Dr. Yingxi Zhu's research group for his help with initial AFM instrumentation.

## Data Availability Statement

The data that support the findings of this study are openly available at https://github.com/cbishop4/Alahe2026.